\documentclass[11pt]{article}
\usepackage{moriond,epsfig}
\def\Journal#1#2#3#4{{#1} {\bf #2}, #3 (#4)}

\def\PRL{\em Phys. Rev. Lett.}
\def\PRD{{\em Phys. Rev.} D}

\def\JHEP{{\em JHEP}}

\def\be{\begin{equation}}
\def\ee{\end{equation}}
\def\bea{\begin{eqnarray}}
\def\eea{\end{eqnarray}}

\begin{document}
\vspace*{4cm}
\title{Jet Reconstruction and Jet Quenching in Heavy Ion Collisions at ATLAS}

\author{ Martin Spousta$^{\ast}$ \\ on behalf of the ATLAS Collaboration }

\address{ $^{\ast}$Charles University in Prague, Institute of Particle and Nuclear Physics, \\ V Holesovickach 2, 180 00 Prague 8, Czech republic}

\maketitle\abstracts{
We present a measurement of dijet asymmetry and dijet azimuthal correlations in Pb+Pb collisions at $\sqrt{s_{NN}} = 2.76$ TeV using the ATLAS detector. This measurement provides the first evidence of a strong jet quenching in relativistic heavy ion collisions at TeV energies. 
The jet reconstruction procedure is discussed as well as studies 
which have been performed to check that the observed asymmetry 
is not produced by detector effects and underlying event
backgrounds.
%The measurement has been published in \ref{thePRL}. 
%In these proceedings, we provide additional supporting material to the original observation reported in December %2010~\cite{thePRL}.
}

\section{Introduction}

%\linenumbers

Ultra-relativistic heavy ion collisions are expected to produce hot and dense QCD matter. One of the main tools to study the production of such matter and its properties is a measurement of jets. Fast quarks or gluons produced in hard processes are expected 
to lose energy and/or have their parton shower modified in the medium of high color-charge density \cite{Urs}. This may lead to a modification of jet yields and/or the structure of jets. Such an effect is called ``jet quenching''. The first indirect evidence for jet quenching has been observed by experiments at the Relativistic Heavy Ion Collider (RHIC) by measuring the spectra of fast hadrons or di-hadron azimuthal correlation \cite{Phenix,Star}. Even if there are many phenomenological models aiming to describe the effect of the jet quenching, there is no unique understanding of mechanisms responsible for the in-medium jet modifications. LHC energies provide an opportunity to study fully reconstructed jets and their properties. In these proceedings we present a first observation of a possible jet modification measured using the ATLAS detector \cite{thePRL}. 

For this study, jets are defined using the anti-$k_\mathrm{T}$ clustering algorithm \cite{} with the distance parameter $R=0.4$. 
The inputs to this algorithm are ``towers''
of calorimeter cells of size $\Delta\eta \times \Delta\phi = 0.1 \times 0.1$ with
the cell energies weighted using energy-density-dependent
factors to correct for calorimeter non-compensation and
other energy losses. Jet four-momenta are constructed
by the vectorial addition of cell four-vectors which are assumed to be massless.
The average contribution from the underlying event (UE) to the jet energy is subtracted from each jet candidate.
The estimate of UE contribution is calculated independently for each event as a 
function of longitudinal calorimeter layer in bins of width $\Delta\eta = 0.1$ by averaging the transverse energy over the azimuth outside of jet regions of interest. 
%For each event, we estimate the average
%transverse energy density in each calorimeter layer in bins
%of width $\Delta\eta = 0.1$, and averaged over azimuth.
Jet regions of interest are selected using a ratio of maximum tower energy to mean tower energy inside a jet which is required to be greater than 5. The value of this discriminant cut is based on simulation studies, and the results have been
tested to be stable against variations in this parameter. The efficiency of the jet reconstruction algorithm,
and other event properties, have been studied using
PYTHIA \cite{PYTHIA} jet events superimposed on HIJING
events \cite{HIJING}.

\begin{figure}
\begin{center}
\mbox{\epsfig{file=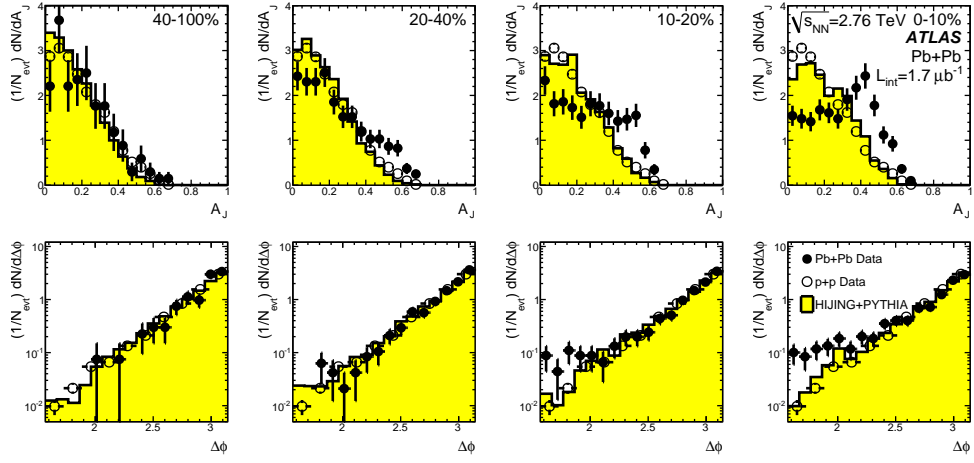,width=6cm,angle=270}}
\caption{\label{Fig:Asym}
(top) Dijet asymmetry distributions for data (points) and unquenched HIJING with superimposed PYTHIA dijets (solid yellow histograms),
as a function of collision centrality (left to right from peripheral
to central events).  Proton-proton data from $\sqrt{s}=7$ TeV, analyzed with the same jet selection, is shown as open circles.
(bottom) Distribution of $\Delta\phi$, the azimuthal angle between the two jets, for data and HIJING+PYTHIA, also as a function of centrality.
}
\end{center}
\end{figure}

\section{Di-jet asymmetry and azimuthal correlation}

The cross-section of dijet production is a dominant contribution to the total jet production cross-section.
Jets are therefore most often produced in pairs well balanced in azimuth and transverse energy. 
Jet quenching may lead to an imbalance in the transverse energy since each jet, or initial parton, traverses a different path length in the QCD medium. Such an imbalance can be quantified using the asymmetry defined as $A_{J} = (E_{T,1} - E_{T,2})/(E_{T,1} + E_{T,2})$, where $E_{T,1} > E_{T,2}$  are transverse energies of jets in a dijet system. 
We focus on the balance between
the highest transverse energy pair of jets in events. These jets
are required to have an azimuthal angle separation, $\Delta \phi = | \phi_1 - \phi_2| > \pi/2$ to reduce contributions from multi-jet
final states.  
%Jets with $\Delta \phi > \pi /2$ can be labeled as being in opposite hemispheres. 
Furthermore the first jet is required to have $E_{T,1}>100$~GeV, and the second jet $E_{T,2}>25$~GeV. 
The jet selection is
chosen such that the first (leading) jet has high reconstruction efficiency and the second (sub-leading) jet is above the distribution of
background 
fluctuations and soft jets associated
with the collision. The jet selection criteria yield a sample of 1693 events from the 2010 Pb+Pb data corresponding to an integrated luminosity of approximately
$1.7~\mu$b$^{-1}$. The dijets are expected to have the asymmetry with a maximum near zero and rapidly decreasing towards the kinematic limit determined by the selected cuts which lies near the asymmetry of 0.7. 

%The shape of the reference
Figure \ref{Fig:Asym} shows the result of the measurement, upper plots show the dijet asymmetry, lower plots show the dijet azimuthal correlations.
The measurement is evaluated in four bins of collision centrality going from the most central (0-10\%) to the most peripheral (40-100\%).	
The centrality is defined using the total sum of transverse energy ($\Sigma E_T$) deposited in the forward calorimeters (FCal).
The asymmetry distribution for dijets measured in p+p collisions at $\sqrt{s} = 7$~TeV is shown in open symbols in the upper plots of Fig.~\ref{Fig:Asym}.   
The yellow distributions show the Monte Carlo (MC) reference which consists of fully reconstructed PYTHIA dijets embedded into the underlying event simulated by the HIJING MC generator. The presence of dijets with large asymmetries both in the reference samples and p+p data
reflects the contribution from events with a topology of three or more jets, and the detector effects. 
Compared to the reference, the asymmetry measured in central heavy ion collisions exhibits the absence of the global maximum near zero and a rather flat plateau for asymmetries below $\approx 0.4$. With decreasing centrality the asymmetry is getting smaller and it returns back to a good agreement with the reference p+p data and MC simulations. Simultaneously with the dijet asymmetry we measure the dijet azimuthal correlations which are presented in the lower plots of Fig.~\ref{Fig:Asym}, again for different centrality bins. The measurement of dijet azimuthal correlations ensures that the effect clearly seen in central collisions does not come from correlated background fluctuations or detector effects. The back-to-back configuration of dijets persists even in the most central collisions.

Numerous studies have been performed to verify that
the events with large asymmetry are not produced by 
backgrounds or detector effects. In the following section we will briefly discuss some of the studies performed to ensure the correctness of the obtained result. 

%The presented measurement has been cross-checked against possible biases coming from: the background subtraction, the jet energy scale shifts, the jet energy resolution %smearing. Left plot of Fig.~\ref{} shows the ... 

\begin{figure}
\begin{center}
\mbox{\epsfig{file=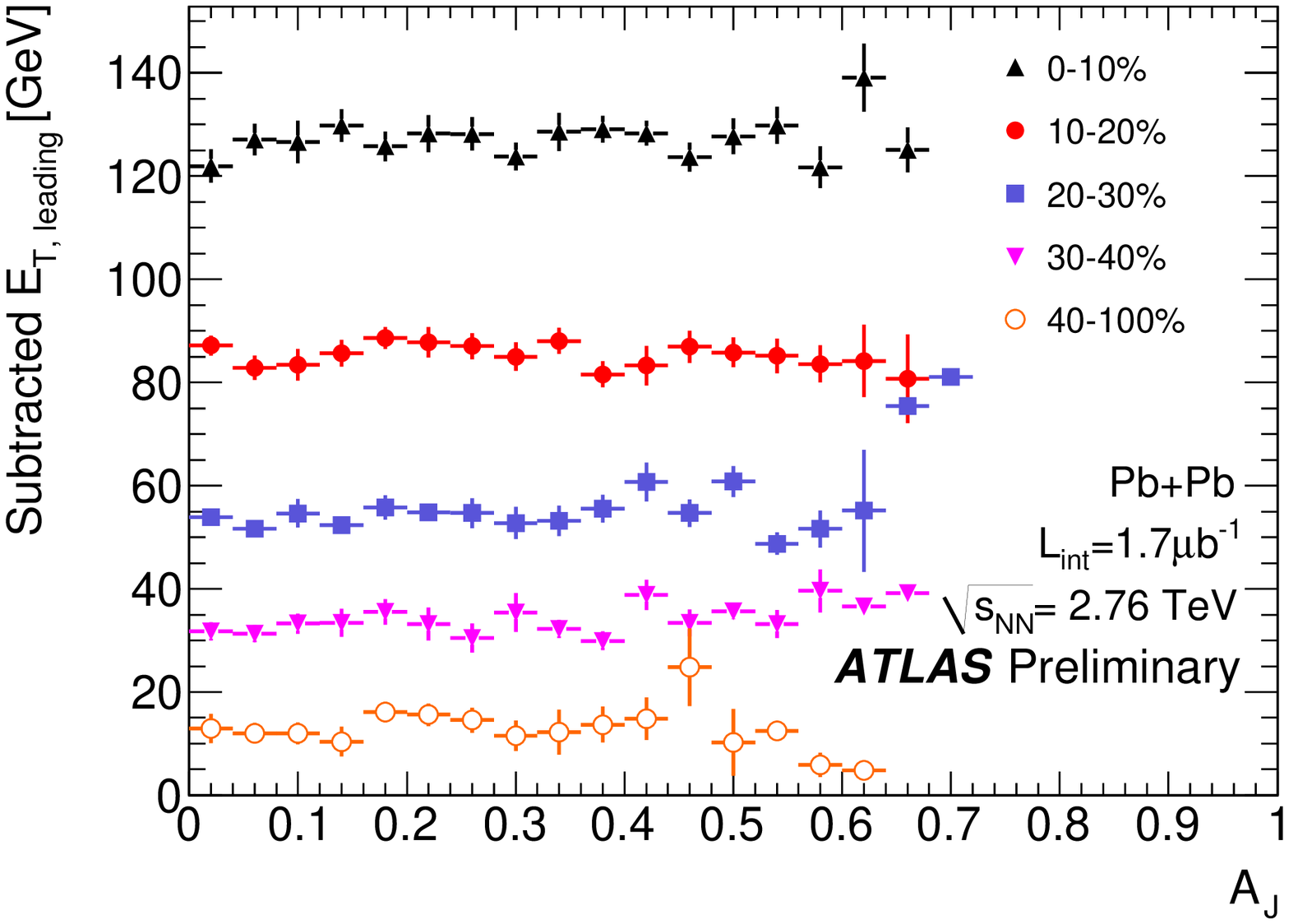,width=6cm}}
\mbox{\epsfig{file=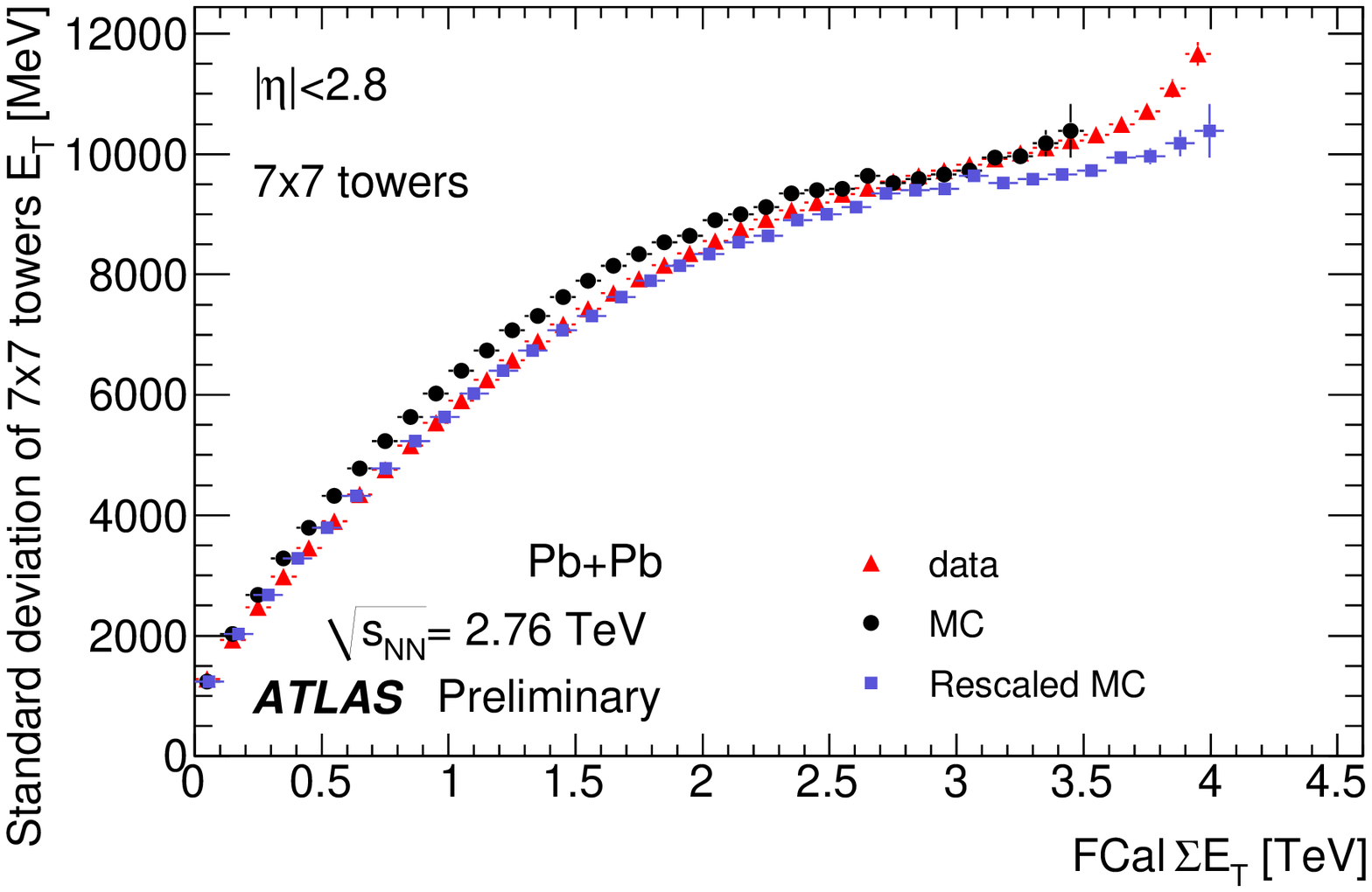,width=6.72cm}}
\end{center}
\caption{\label{Fig:Perf}
(left) Mean value of the UE $E_{T}$ subtracted form the leading jet as a function of dijet asymmetry. (right) Mean standard deviation of transverse energy at electromagnetic scale in $0.7 \times 0.7$ windows ($7 \times 7$ calorimeter towers) as a function centrality quantified using the FCal $\Sigma E_T$. The data (red) are compared to MC with (blue) and without (black) rescaling of the endpoint of the FCal $\Sigma E_T$ MC distribution to the endpoint measured in data. The rescaling allows a direct comparison with the data which extend to larger FCal $\Sigma E_T$ than the MC.
}
\end{figure}

\section{Further cross-checks and measurement of the energy flow}

One of the possible biases might come from the subtraction of the underlying event during the jet reconstruction. To check that the underlying event subtraction does not influence the measured asymmetry distribution we evaluate the amount of subtracted energy as a function of the dijet asymmetry. The result for the leading jet is shown in the left plot of Fig.~\ref{Fig:Perf}. The size of the background subtraction does not change for jets with large asymmetry, neither for the leading nor for the sub-leading jet. 
%To assure that the asymmetry is not produced by the jet energy resolution smearing in central events we 
To further test the reliability of the comparison of data with HIJING %, and in particular to test the appropriate modeling of the UE fluctuations, 
we compare the magnitude of the calorimeter fluctuations in the Minimum Bias reconstructed HIJING events with those in Pb+Pb events. The calorimeter fluctuations are quantified by the mean standard deviation of the non-calibrated transverse energy of towers grouped in ``windows'' of $7 \times 7$ towers. The $7 \times 7$ window approaches the size of an average jet defined using the anti-$k_T$ algorithm with $R=0.4$. The standard deviation is calculated event by event and the mean is evaluated in fine bins of centrality. The right plot of Fig.~\ref{Fig:Perf} shows the comparison of the mean standard deviation as a function of centrality for MC and data. As before, the centrality is defined using the total sum of transverse energy deposited in forward calorimeters. The 0-10\% central events correspond to FCal $\Sigma E_T$ greater than approximately 2.4 TeV. 
One can see a very good correspondence between the data and MC suggesting a good modeling of UE fluctuations and therefore appropriate modeling of the jet energy resolution.

\begin{figure}
\begin{center}
\mbox{\epsfig{file=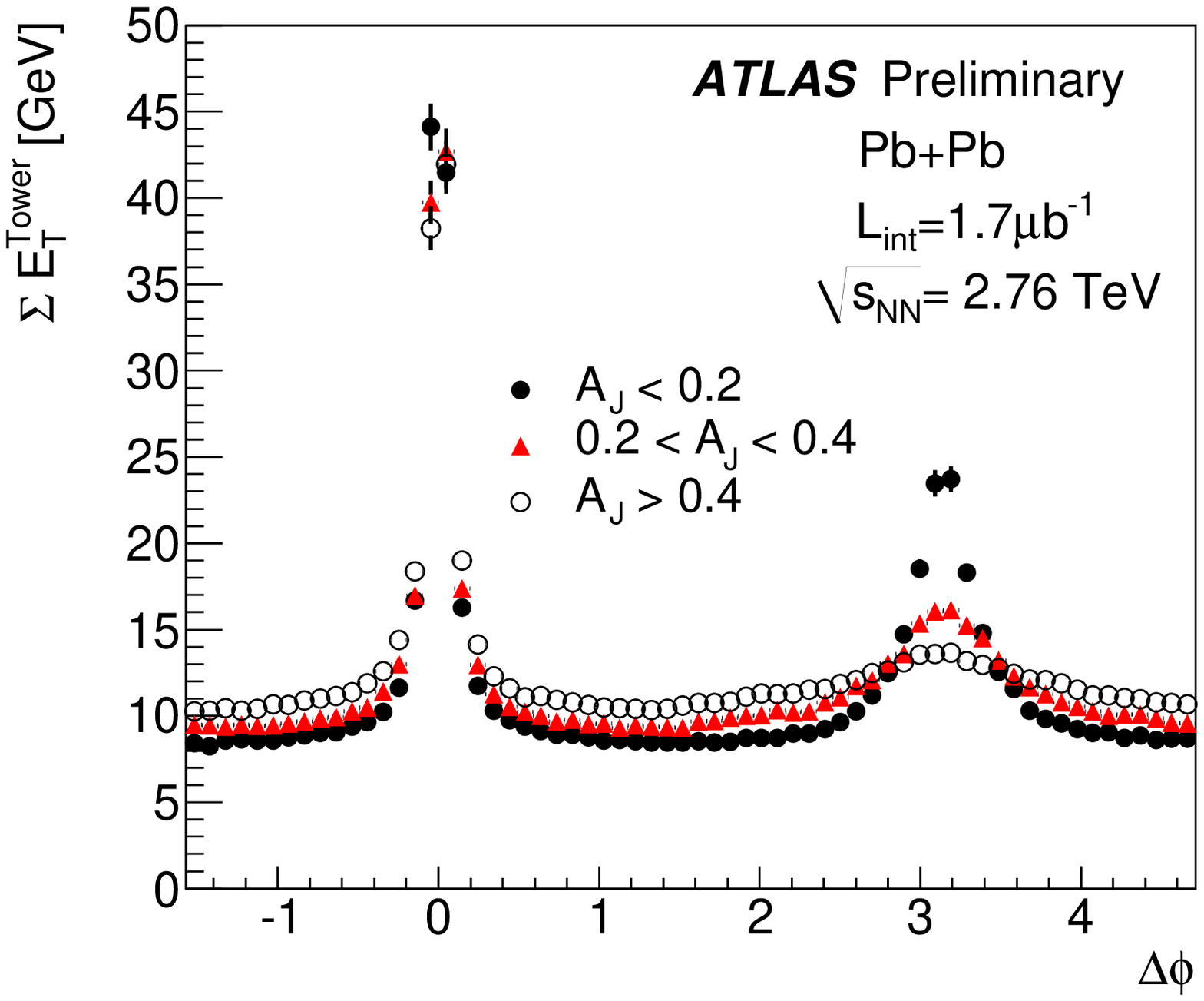,width=6cm}}
\mbox{\epsfig{file=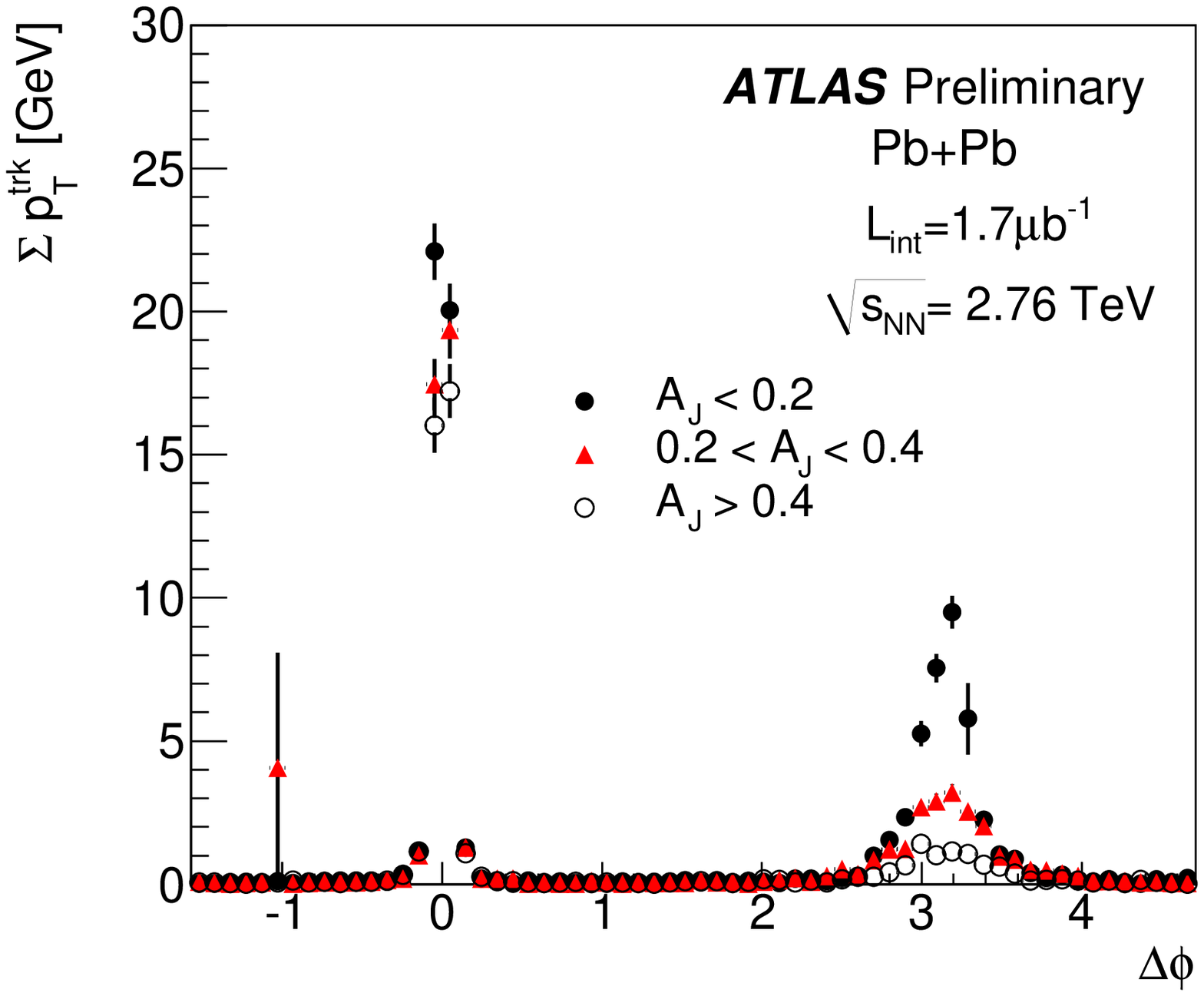,width=6cm}}
\end{center}
\caption{\label{Fig:Flow}
Energy flow measured using non-calibrated non-subtracted calorimeter towers (left) and tracks with $p_{T} > 4$~GeV (right).
The transverse energy of towers or transverse momentum of tracks is summed over strips of size, $\Delta\eta \times \Delta\phi = 0.8 \times 0.1$, centered at the pseudorapidity position of the leading jet or sub-leading jet.
}
\end{figure}

One step further in understanding of the origin of the large dijet asymmetry is a measurement of the energy flow or momentum flow in the event. In this measurement, the
transverse energy is summed over strips of size $\Delta\eta \times \Delta\phi = 0.8 \times 0.1$, centered at the pseudorapidity position of the leading jet or sub-leading jet. % delta phi < \pi/2 ...
The
dependence of the energy on the
azimuthal angle measured with respect to the leading jet strip 
provides a
method for evaluating the jet
characteristics without requiring 
per-event background subtraction. The transverse energy sum is also independent of the jet calibration which provides a further cross-check that the measurement is not an artifact of a bad calibration.
The left plot of Fig.~\ref{Fig:Flow} shows the sum of the transverse energy for three bins in the measured jet asymmetry. 
The asymmetry is clearly visible even at the level of non-calibrated, non-subtracted towers.
One can also see the overall offset of the distribution due to the UE event which is not subtracted. 
The offset of the distribution is larger for jets with larger asymmetries, since these occur in more central collisions. The similar measurement of the energy flow is performed using the charged particles measured in the Inner Detector as shown in the right plot of Fig~\ref{Fig:Flow}. The threshold on minimum $p_T$ of charged particles was selected to be 4 GeV in order to suppress tracks coming from UE. The result for charged particles is similar to that obtained from calorimeter towers. The offset due to the UE is not visible since the 4 GeV cut effectively suppresses the particles coming from the underlying event.

\section{Conclusions}

We observe a large dijet asymmetry in a sample of events with a reconstructed
jet with transverse energy of 100 GeV or more.
The asymmetry that is observed between the transverse energies of the
leading and sub-leading jets increases with the centrality
of the collisions. The measured dijets remain well correlated in azimuth. 
The result has been tested to exclude possible biases such as detector effects, 
jet energy scale and resolution, and background subtraction. 
The natural interpretation of the observation is a strong jet quenching present in central heavy ion collisions.


\section*{References}
\begin{thebibliography}{99}

\bibitem{Urs} U. A. Wiedemann, arXiv:0908.2306 
\bibitem{Phenix} PHENIX Collaboration, K. Adcox et al., \Journal{\PRL}{88}{022301}{2002}%Phys. Rev. Lett. 88 (2002) 022301
\bibitem{Star} STAR Collaboration, C. Adler et al., \Journal{\PRL}{90}{082302}{2003}
%Phys. Rev. Lett. 89, 202301 (2002)
\bibitem{thePRL} ATLAS Collaboration, \Journal{\PRL} {105}{252303}{2010}
\bibitem{PYTHIA} T. Sjostrand, S. Mrenna, P. Z. Skands \Journal{\JHEP}{0605}{026}{2006}
\bibitem{HIJING} X.-N. Wang, M. Gyulassy, \Journal{\PRD}{44}{3501}{1991}
\end{thebibliography}
\end{document}